\documentclass[a4paper,11pt]{article}
\pdfoutput=1 

\usepackage{jinstpub} 
\usepackage{verbatim}
\usepackage{siunitx}

\usepackage{amsmath}
\newcommand*\subtxt[1]{_{\textnormal{#1}}}
\DeclareRobustCommand\_{\ifmmode\expandafter\subtxt\else\textunderscore\fi}

\usepackage{caption} 
\newcommand{\Cd}{\mathrm{^{109}Cd}}

\newcommand{\ipreamp}{I\_{\it preamp}} 
\newcommand{\sigtoa}{$\sigma\_{\it TOA0-TOA1}~$} 
 
\title{\boldmath  Efficiency and time resolution of monolithic silicon pixel detectors in SiGe BiCMOS technology}

\author[a,1]{G. Iacobucci,\note{Corresponding author.}}
\author[a,b]{L. Paolozzi,}
\author[a]{P. Valerio,}
\author[a]{T. Moretti,}
\author[a]{F. Cadoux,}
\author[a,2]{R. Cardarelli\note{Also at INFN Section of Roma Tor Vergata, Via della ricerca scientifica 1, Roma, Italy.},}
\author[a]{R. Cardella,}
\author[a]{S. Débieux,}
\author[a]{Y. Favre,}
\author[a]{D. Ferrere,}
\author[a]{S. Gonzalez-Sevilla,}
\author[a]{Y. Gurimskaya,}
\author[a,b]{R. Kotitsa,}
\author[a]{C. Magliocca,}
\author[b,c]{F. Martinelli,}
\author[a]{M. Milanesio,}
\author[a]{M. Münker,}
\author[a,b]{M. Nessi,}
\author[a,b]{A. Picardi,}
\author[a]{J. Saidi,}
\author[d]{H. Rücker,}
\author[a]{M. Vicente Barreto Pinto,}
\author[b]{S. Zambito}

\affiliation[a]{D\'epartement de Physique Nucl\'eaire et Corpusculaire (DPNC),
University of Geneva, 24 Quai Ernest-Ansermet, CH-1211 Geneva 4, Switzerland}

\affiliation[b]{CERN, CH-1211 Geneva 23, Switzerland}

\affiliation[c]{École polytechnique fédérale de Lausanne (EPFL), Advanced Quantum Architecture (AQUA) Laboratory Rue de la Maladière 71C, 2002 Neuchâtel, Switzerland}

\affiliation[d]{IHP — Leibniz-Institut für innovative Mikroelektronik, Im Technologiepark 25, Frankfurt (Oder), Germany}

\emailAdd{giuseppe.iacobucci@unige.ch}

\abstract{A monolithic silicon pixel detector prototype has been produced in the SiGe BiCMOS SG13G2 130 nm node technology by IHP. The ASIC contains a matrix of hexagonal pixels with  pitch of approximately 100 $\mu$m. Three analog pixels were calibrated in laboratory with radioactive sources and tested in a 180 GeV/c pion beamline at the CERN SPS. 
A detection efficiency  of 
(99.9{\raisebox{0.5ex}{\tiny$\substack{+0.1 \\ -0.2}$}})\%
was measured together with a time resolution of $(36.4 \pm 0.8) ~\si{\pico\second}$ at the highest preamplifier bias current working point of 150 $\mu$A and at a sensor bias voltage of 160 V. The ASIC was also characterized at lower bias voltage and preamplifier current.}

\keywords{Particle tracking detectors (Solid-state detectors); Solid state detectors; Instrumentation and methods for time-of-flight (TOF) spectroscopy; Pixelated detectors and associated VLSI electronics}

\begin{document}
\maketitle 
\flushbottom

\section{Introduction}
\label{sec:intro}
Since several years the particle-physics community has devoted an increasing interest to   the timing performance of silicon \cite{werner,Sadrozinski} for the detection of ionising radiation. 
In a pixelated hybrid detector, the NA62 GigaTracker \cite{NA62GT} achieved time resolution down to 150 ps. 
The ATLAS and CMS Collaborations are upgrading their detectors for the High Luminosity LHC program  with planes of  Low Gain Avalanche Detectors (LGAD) \cite{atlasTDR,cmsTDR}.  
LGADs are planar silicon sensors that exploit the increase of the signal-to-noise ratio given by an internal gain of 10-50 to provide time resolutions of approximately 30 ps. Those employed in the ATLAS and CMS timing layer have an area of about 1.6 mm$^2$.

An alternative approach to obtain the signal-to-noise ratio necessary to achieve such time resolution comes from the exploitation of the unparalleled analog performance of Silicon-Germanium Heterojunction Bipolar Transistors (SiGe HBT) to produce a fast signal amplification with low noise at low amplifier current density \cite{Cressler2013}\cite{Mai2016SiGeBiCMOSBT}. An R$\&$D effort started in recent years to exploit commercial SiGe BiCMOS - a mainstream microelectronic technology with Very Large Scale Integration (VLSI) capability - to produce monolithic pixel detectors with small pixel size and time resolution comparable with that of LGAD, without recurring to an avalanche gain mechanism \cite{discrete_SiGe}\cite{Valerio_2019}.

Laboratory measurements with
radioactive sources of the first monolithic silicon pixel detector prototype in the 130 nm SG13G2 IHP process \cite{SG13G2} with 100 $\mu$m pixel pitch demonstrated that time resolutions at the level of 50 ps can be achieved \cite{hexa_50ps}\cite{Paolozzi_2020}.
Such remarkable timing performance for a silicon sensor without an internal gain layer was obtained at rather high discrimination threshold, which was possibly affecting the sensor detection efficiency. 
Furthermore, 
there were indications
that  the resolution on the signal time-over-threshold (TOT), which was used to correct for signal time-walk, was limiting the timing performance of the detector. These considerations suggested that a study of the analog signals and an improved  time-walk correction method would be necessary to fully exploit the timing capability of this type of silicon detector maintaining at the same time very high efficiency.

For this purpose a second monolithic silicon pixel ASIC prototype was produced in the same 130 nm IHP process. The ASIC contains four sub-matrices of pixels with digital readout as well as four analog pixels with the preamplification stage followed by an analog driver with dedicated output pads.
Although a full description of the ASIC  is given in Section~\ref{sec:chip}, the measurements presented in this paper focus on the analog pixels, since the possibility to measure the full output waveform allows for a better insight on the circuit performance, decoupling the front-end characteristics from the digital logic, therefore permitting a detailed study of the efficiency and timing response.
The analog channels were characterized in  laboratory tests with radioactive sources at the University of Geneva and with minimum ionizing particles (MIPs) at the CERN SPS  testbeam facility.

\section{Description of the ASIC}
\label{sec:chip}

\subsection{Architecture and Floorplan}

\begin{figure}[ht]
\centering
\includegraphics[width=.55\textwidth,trim=0 0 0 0]{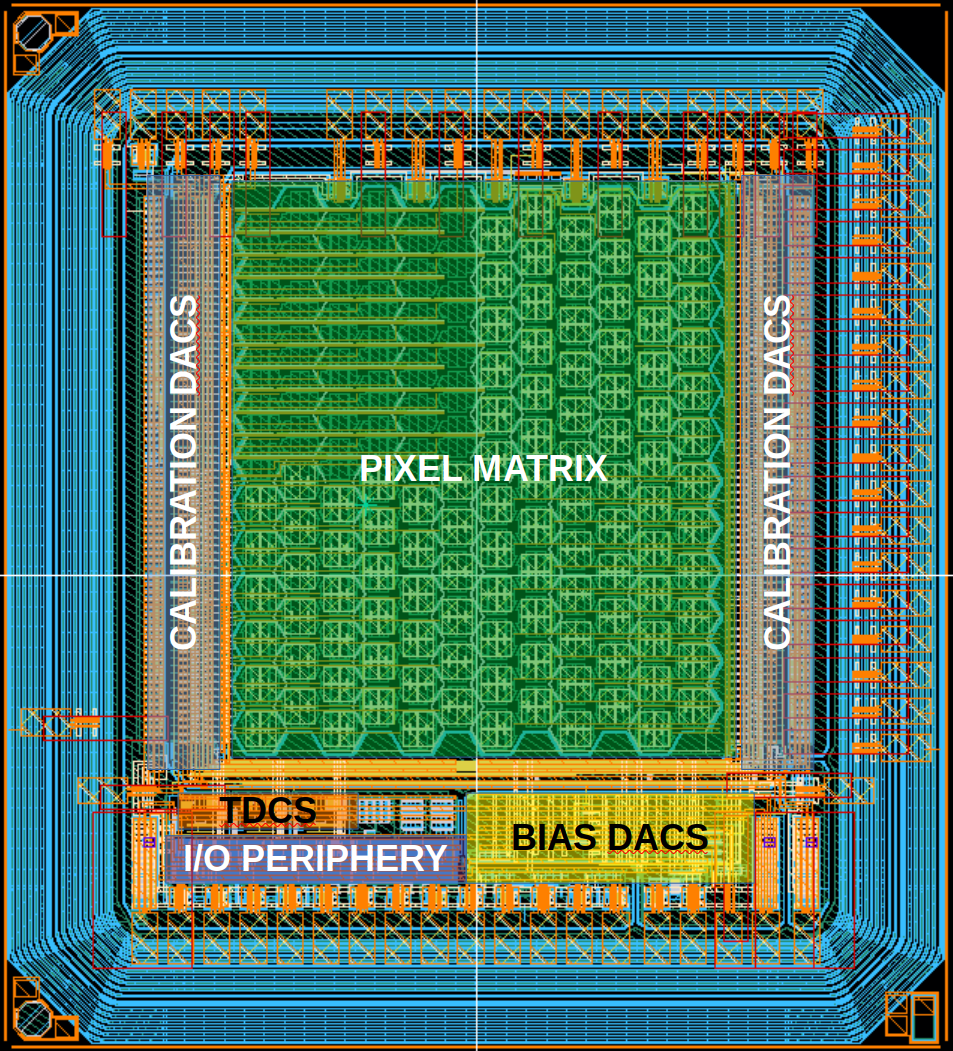}
\caption{\label{fig:chip_layout} Floorplan of the  monolithic pixel detector prototype. The ASIC size is 2.3$\times$2.5 mm$^2$.
It contains four sub-matrices of 6$\times$6 hexagonal pixels with a pitch of approximately 100~$\mu$m.}
\end{figure}

The ASIC prototype was produced in the IHP SG13G2 \SI{130}{\nano\meter} BiCMOS technology with a wafer resistivity of \SI{50}{\ohm\centi\meter}. It features a matrix of 12$\times$12 monolithic hexagonal pixels of 65 $\mu$m side with a pixel capacitance of 80 fF\footnote{The sensor implemented in the pixel is essentially the same sensor of \cite{Paolozzi_2020}.}.
The ASIC floorplan is shown in Figure \ref{fig:chip_layout}.
Each pixel includes an HBT-based frontend, a fast discriminator and an 8-bit threshold-tuning DAC.
A common digital logic placed beside the matrix (from now on called "periphery") contains the readout logic, a dedicated timing digitization circuit, a number of DACs to provide the bias voltages, and currents to the pixels and an SPI slow-control interface to program them.

The pixels are built using the n-well as sensor cathode, biased at a positive bias voltage that can be changed  via an external power supply, and the p-doped substrate as the anode. The anode is biased at a negative high voltage ($\le$ \SI{-80}{\volt} in the data presented here). The presence of this high-voltage bias forces all the electronics to be insulated by the substrate it in order to avoid breakdown. Hence, every nMOS is placed in insulated p-wells, divided in three power domains. A series of 18 guard-rings are placed around the edge of the chip to insulate the substrate.

The output of the discriminator is  sent to a Time-to-Digital Converter (TDC) positioned in the periphery to calculate a timestamp for the hit and measure the TOT of the signal.
The ASIC integrates three separate TDCs, each wired to a fast-or line connected to different pixels. The fast-or lines are interleaved so that two adjacent pixels are always connected to separate TDCs: this ensures that in the event of a particle hitting multiple adjacent pixels, the timing information can be measured from all of them independently.
The TDC architecture is described in \cite{fulvio_tdc}. A 9-stage pseudo-nMOS feed-forward ring oscillator is used with a set of latches to sample the oscillator state. The TDC resolution is about \SI{30}{\pico\second}.
Each TDC features three measurement channels: one for the event timestamp, one for ToT and one for calibration purposes.

\subsection{Front-end flavors}
\begin{figure}[ht]
\centering
\includegraphics[width=.7\textwidth,trim=0 0 0 0]{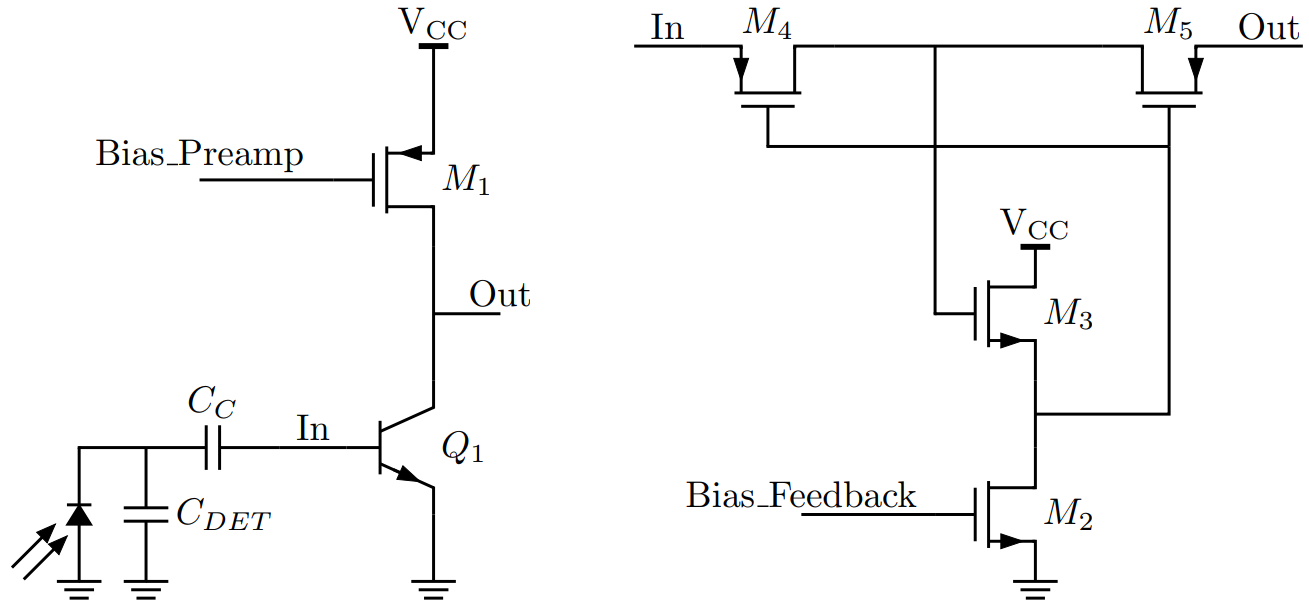}
\caption{\label{fig:preamp} Schematic layout of the BJT-based preamplifier of the  prototype ASIC. The left block is a common-emitter configuration capacitively coupled to the sensor, while the right one implements a floating MOS-based feedback resistor.
}
\end{figure}

The pixel matrix is divided into four 6$\times$6 sub-matrices, each with a different amplifier design. The discriminator, calibration DACs and readout interface are the same. The basic schematic of the preamplifier can be seen in Figure \ref{fig:preamp}. It consists of a SiGe HBT used as a charge sensitive amplifier, with a pMOS load and an MOS-based floating resistor used as a feedback. The architecture is similar to the one the authors developed for \cite{hexa_50ps}, which yielded to excellent 
timing performance.
This circuit is followed by a MOS discriminator that compares the output of the preamplifier to a fixed threshold to determine if the pixel was hit by a particle.
The four sub-matrices contain different variations of this circuit:

\vspace{-9pt}
\begin{itemize}
\setlength\itemsep{-4pt}
    \item The circuit in Figure \ref{fig:preamp}, with the preamplifier, discriminator and calibration DAC placed outside of the active pixel area;
    \item The same circuit, with the preamplifier integrated in the pixel n-well;
    \item The same amplifier, but with an added HBT common collector driving stage placed before the discriminator;
    \item A version of the circuit with two discriminators with different thresholds,  to calculate the rising slope of the input signal directly
    and perform a more accurate time-walk correction.
\end{itemize}

\vspace{-9pt}

\subsection{Analog pixels}

\begin{figure}[ht]
\centering
\includegraphics[width=.6\textwidth,trim=0 0 0 0]{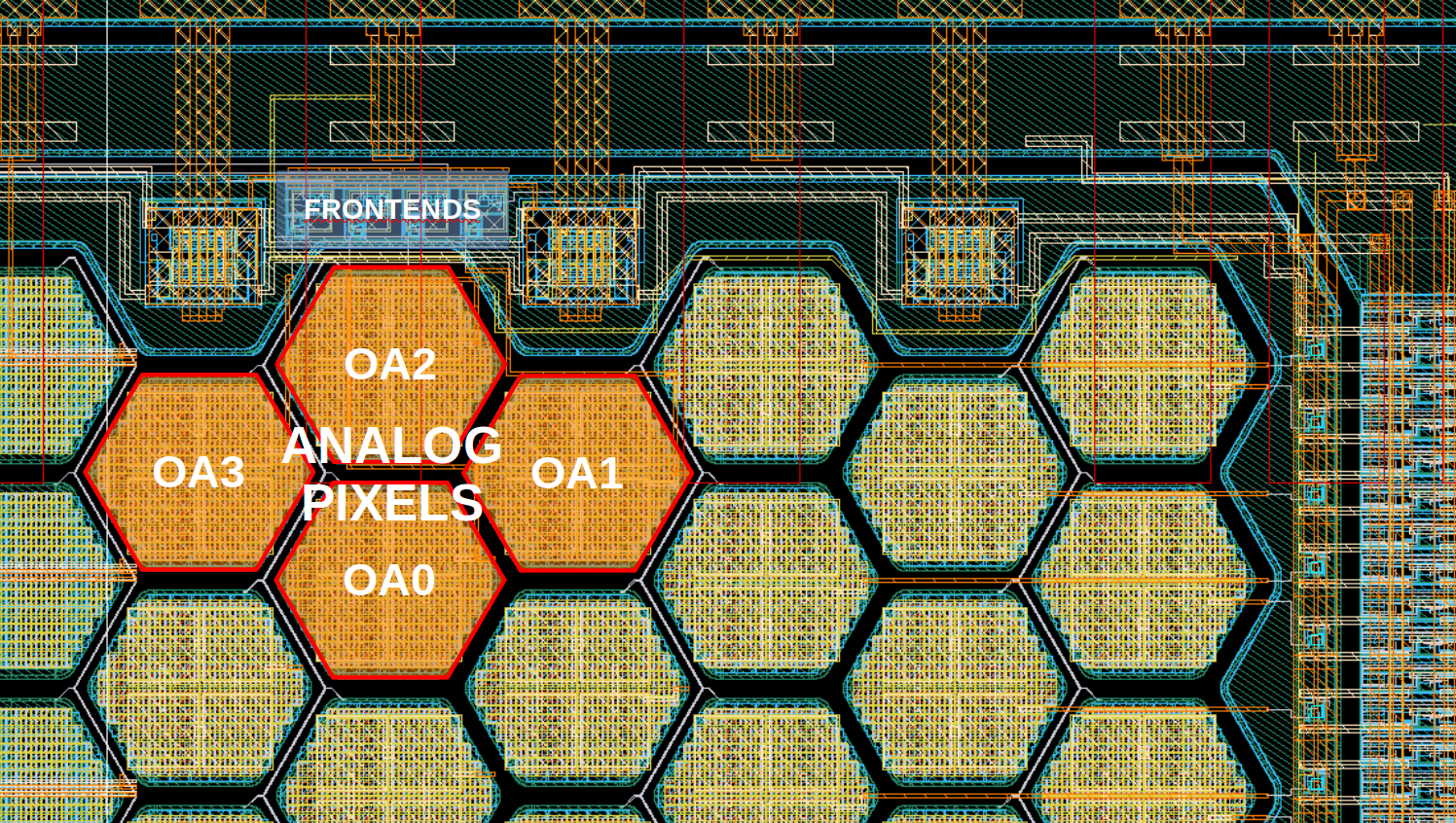}
\caption{\label{fig:analog_pixels} Detail of the ASIC layout that shows the four pixels directly connected to output pads. During the CERN SPS testbeam, pixels OA0, OA1 and OA2 were read by an oscilloscope. This study is based on the data acquired with these three analog pixels.}
\end{figure}

One of the sub-matrices hosts four analog pixels (Figure \ref{fig:analog_pixels}). They contain the same HBT amplifier represented in Figure \ref{fig:preamp}, followed by an analog driver  directly connected to output pads to be measured with an oscilloscope. The driver is composed by two consecutive HBTs in common collector configuration, with the first stage AC coupled to the output of the amplifier. The first stage of the driver has a configurable output impedance, while the second stage is terminated on a \SI{500}{\ohm} resistor in chip. The entire front-end is placed outside of the pixel area, in proximity of the output pads, as shown in Figure \ref{fig:analog_pixels}. The short routing from the pixel to the amplifier input is estimated to contribute a capacitance of less than 10 fF.

\subsection{Slow-control and readout interface}

The periphery of the chip contains, in addition to the I/O pads, thirteen 8-bit DACs to tune all the bias voltages and currents used by the analog blocks in the pixels. Each DAC is programmable via a standard SPI interface using 16-bit commands (8 to address a specific DAC, 8 for the DAC value). The same SPI interface can be used to program a "mask" bit for individual pixels (i.e. force their output to be ignored by the readout logic). This is useful in case of malfunctioning pixels, or to test only some parts of the chip at a time.
A special SPI command is also used to initiate a frame readout. As soon as this command is sent, the content of the TDCs and the address of the pixel hit is read out via a low-voltage differential line. The state of the pixel is then reset, to be able to acquire a new frame.

\section{Amplifier response calibrations}
\label{sec:calibrations}

Laboratory measurements with X-ray sources were performed to characterize the analog response of the front-end electronics of two ASICs, that will be called here Detector Under Study 0 (DUT0) and 1 (DUT1). For these measurements and those described in Section \ref{sec:results}, the outputs of the analog drivers were connected to a \SI{50}{\ohm} coaxial cable and AC coupled via a \SI{100}{\nano\farad} capacitor to a Lecroy WaveMaster 820zi oscilloscope. The oscilloscope has a sampling rate of 40 Gs/s and   analog bandwidth limited to 4 GHz. 
A $ \Cd $ source was used to measure the gain of the electronic chain, which comprises the pixel, the preamplifier and the driver. To study the response of the electronics as a function of the power consumption, four different working points were adopted, with  amplifier bias current $\ipreamp = $  \SI{7}{\micro\ampere}, \SI{20}{\micro\ampere}, \SI{50}{\micro\ampere} and \SI{150}{\micro\ampere}.

\begin{figure}[!htb]
	\centering
	\includegraphics[width=.99\textwidth,trim=0 0 0 0, angle =0 ]{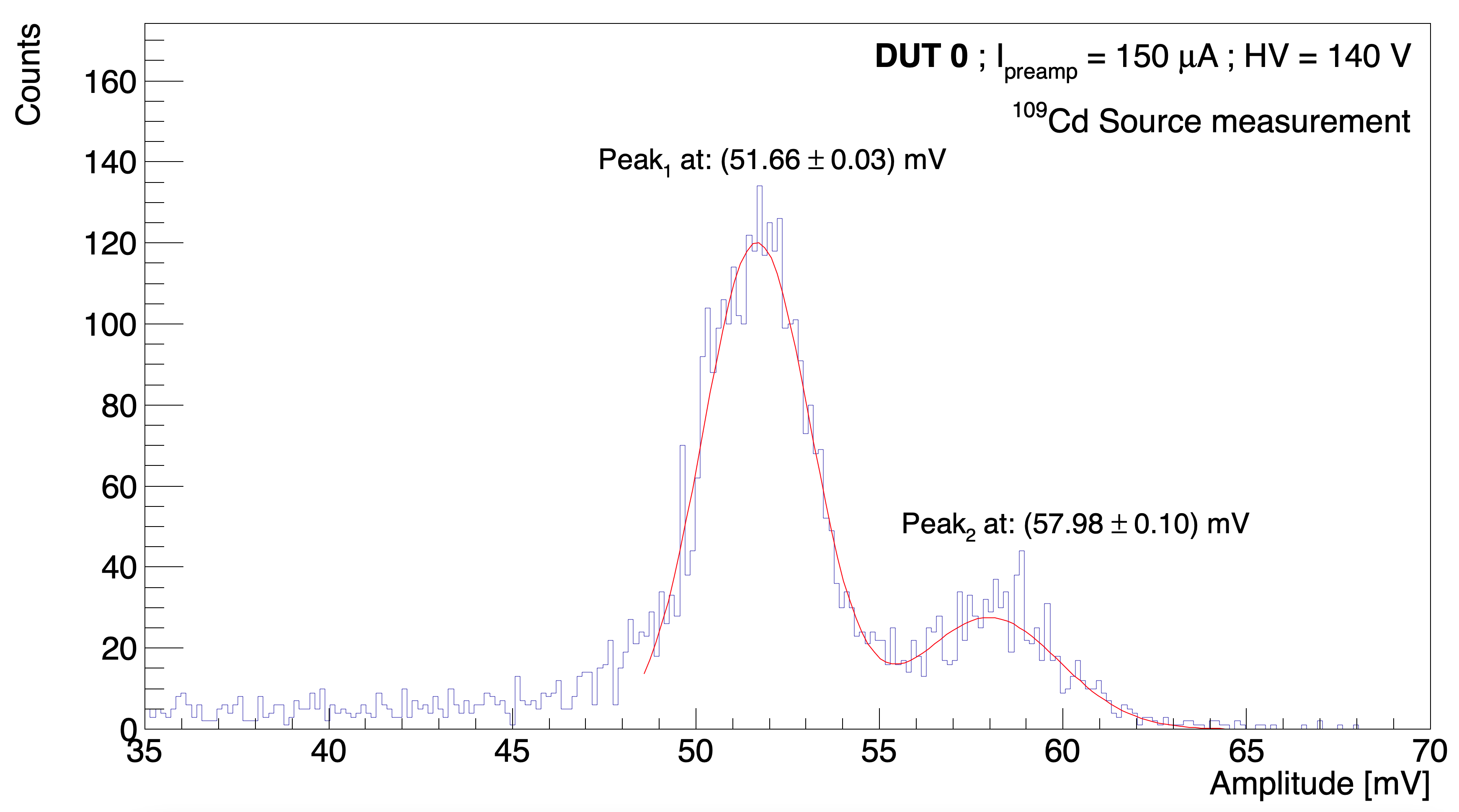}
	\caption{Spectrum of the $ \Cd $ source measured at the working point  $\ipreamp = $\SI{150}{\micro\ampere} and $HV=140~V$. The two peaks, corresponding to the X-ray emission of $ ^{109}Ag $,  were fitted with a double Gaussian function.
}
	\label{fig:CdSpectrum}
\end{figure}

\begin{figure}[!htb]
	\centering
	\includegraphics[width=.72\textwidth, trim=0 0 0 70, clip]{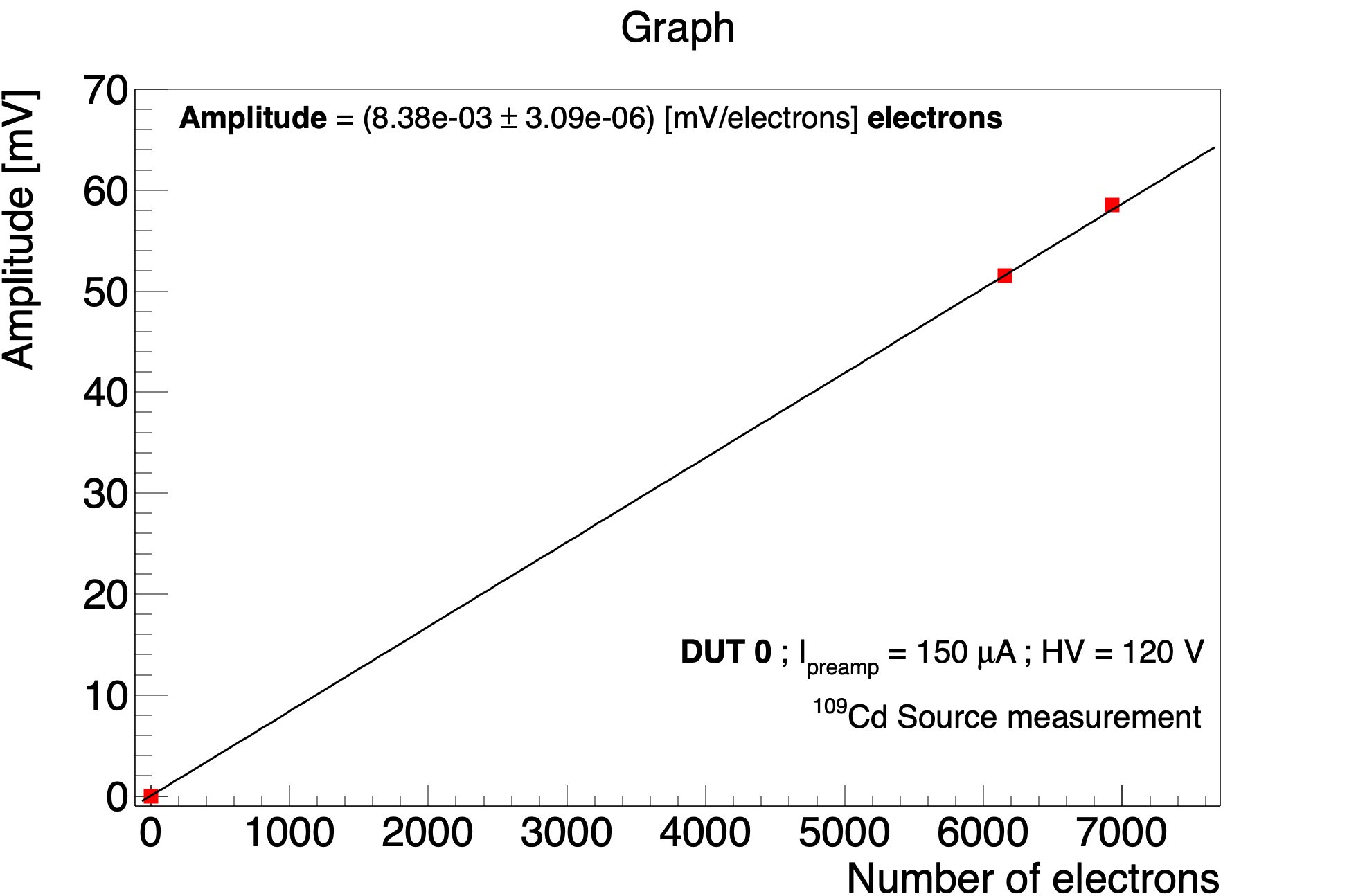}
	\caption{Example of the amplitude measurement as a function of the number of electrons obtained with the $ \Cd $ source for one of the working points for DUT0. The superimposed linear fit  is used to parameterise  the response of the front-end at this working point.}
	\label{fig:CdFit}
\end{figure}

The two main photons from the $ \Cd $ source derive from the $ ^{109}Ag $ X-ray K lines of energy
$E_{1} = \SI{22.16}{\kilo\electronvolt}$
and
$E_{2} = \SI{24.94}{\kilo\electronvolt} $.
The spectrum of $ \Cd $ was fitted with a double Gaussian function. Figure \ref{fig:CdSpectrum} shows one example of the spectra from DUT0, while Figure \ref{fig:CdFit} shows the amplitude values obtained from the fit of the same spectrum as a function of the deposited charge, assuming an energy of \SI{3.6}{\electronvolt} to generate and electron-hole pair in silicon. A linear fit of the data was used to estimate the charge gain ($ A_q $) of the front-end electronics. 

The Equivalent Noise Charge (ENC) of the front-end was then calculated as

\vspace{-3mm}
\begin{equation*}
ENC = \frac{\sqrt{\sigma_V^2-\sigma\_{\it scope}^2}}{A_q}
\end{equation*}

\vspace{-1mm}
where $\sigma_V$ is the 
standard deviation of the electronics-noise distribution 
measured on the oscilloscope when the front-end is connected, and $\sigma\_{\it scope} = \SI{420}{\micro\volt}$ is the voltage noise of the oscilloscope with the open input connector. The results of the gain and ENC measurements are summarized in Table 
\ref{tab:gaintable}. The two DUTs show the same trend within approximately 10\%. At $ \ipreamp = \SI{50}{\micro\ampere} $ the ENC is as low as 50 electrons. As expected, it increases significantly for smaller $ \ipreamp $ values, which could produce a degradation of the detection efficiency. The increase of the ENC at $ \ipreamp = $ 150 µA is due to the configuration of the amplifier used for this working point.

\begin{table}[h!]
\centering
\begin{tabular}{cc|ccc|ccc|}
\cline{3-8}
 &     & \multicolumn{3}{c|}{\textbf{DUT0}}  & \multicolumn{3}{c|}{\textbf{DUT1}} \\ \hline
\multicolumn{1}{|c|}{\begin{tabular}[c]{@{}c@{}}$\ipreamp$\\ {[}$\mu$A{]}\end{tabular}} & \begin{tabular}[c]{@{}c@{}}Power density\\ {[}mW/cm$^2${]}\end{tabular} & \multicolumn{1}{c|}{\begin{tabular}[c]{@{}c@{}}$\sigma_V$\\ {[}mV{]}\end{tabular}} & \multicolumn{1}{c|}{\begin{tabular}[c]{@{}c@{}}A$_q$\\ {[}mV/fC{]}\end{tabular}} & \begin{tabular}[c]{@{}c@{}}ENC\\ {[}electrons{]}\end{tabular} & \multicolumn{1}{c|}{\begin{tabular}[c]{@{}c@{}}$\sigma_V$\\ {[}mV{]}\end{tabular}} & \multicolumn{1}{c|}{\begin{tabular}[c]{@{}c@{}}A$_q$\\ {[}mV/fC{]}\end{tabular}} & \begin{tabular}[c]{@{}c@{}}ENC\\ {[}electrons{]}\end{tabular} \\ \hline
\multicolumn{1}{|c|}{7}                                                                   & 84                                                                      & \multicolumn{1}{c|}{1.35}                                                          & \multicolumn{1}{c|}{42.3}                                                        & 194 $\pm$ 1                                                 & \multicolumn{1}{c|}{1.41}                                                          & \multicolumn{1}{c|}{49.2}                                                        & 171 $\pm$  2                                                  \\ \hline
\multicolumn{1}{|c|}{20}                                                                  & 240                                                                     & \multicolumn{1}{c|}{0.82}                                                          & \multicolumn{1}{c|}{43.4}                                                        & 102 $\pm$ 1                                                   & \multicolumn{1}{c|}{0.83}                                                          & \multicolumn{1}{c|}{46.0}                                                        & ~97 $\pm$ 1                                                    \\ \hline
\multicolumn{1}{|c|}{50}                                                                  & 600                                                                     & \multicolumn{1}{c|}{0.70}                                                          & \multicolumn{1}{c|}{55.0}                                                        & ~54 $\pm$ 1                                                    & \multicolumn{1}{c|}{0.58}                                                          & \multicolumn{1}{c|}{50.5}                                                        & ~50 $\pm$ 1                                                    \\ \hline
\multicolumn{1}{|c|}{150}                                                                 & 1800                                                                    & \multicolumn{1}{c|}{0.81}                                                          & \multicolumn{1}{c|}{52.4}                                                        & ~76 $\pm$ 1                                                    & \multicolumn{1}{c|}{0.72}                                                          & \multicolumn{1}{c|}{44.4}                                                        & ~82 $\pm$ 1                                                    \\ \hline
\end{tabular}
\caption{Power density, voltage noise, charge gain and ENC of the analog front-end electronics at the four $\ipreamp$   working points for HV = 120 V. The power density is calculated for a matrix of pixels with 100 µm pitch. The lower ENC value measured at $\ipreamp = \SI{50}{\micro\ampere}$ is due to a different configuration of the amplifier and driver working point that acts as a filter for noise, thus reducing the bandwidth.}
\label{tab:gaintable}
\end{table}

\section{Efficiency and time resolution measurement}
\label{sec:results}

\subsection{Testbeam experiment setup and data set}
\label{subsec:dataset}

\begin{figure}[!htb]
\centering
\includegraphics[width=.75\textwidth,trim=0 0 0 0]{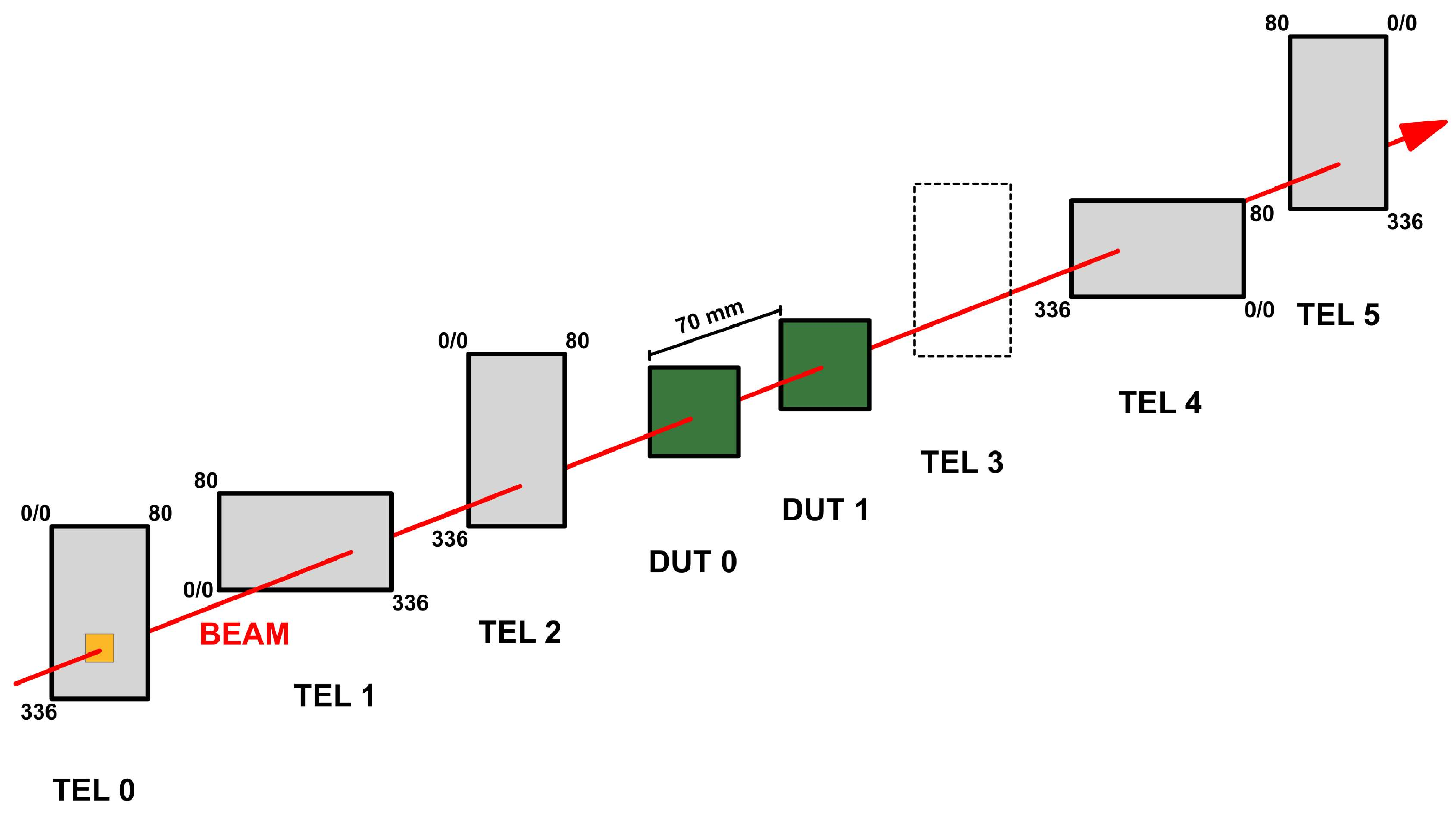}
\caption{\label{fig:Setup} Schematic view of the experimental setup, showing the five FEI4 telescope \cite{Benoit_2016} planes that were operated and the two DUTs in green. Plane TEL3 was not operational during this testbeam measurement. The FEI4 readout chip has a a matrix of $ 80 \times 336 $ pixels with a pixel size of $ 250 \times 50 ~\si{\micro\meter\squared} $. The telescope planes are alternatively rotated by 90$^\circ$  to optimise the space resolution on the two transversal directions. A region of interest, shown by the  yellow area, was imposed to the first plane of the telescope, and was put in coincidence with the last telescope plane to generate the trigger.}
\end{figure}

The detection efficiency and time resolution of the prototypes were measured at the CERN SPS testbeam facility with a pion beam with \SI{180}{\giga\electronvolt}/c momentum. The experimental setup consisted of the UniGe FEI4 telescope for particle tracking \cite{Benoit_2016}, with the two devices under test (DUT0 upstream with respect to the beam, DUT1 downstream) placed after three detection planes of the telescope as shown in Figure \ref{fig:Setup}. The DUTs were operated at room temperature. The DUTs were read by two oscilloscopes with analog bandwidth of 4 GHz and a sampling rate of 40 GS/s and 20 GS/s, respectively. The DUTs were positioned specularly to each other and pixels OA0 (shown in Figure~\ref{fig:analog_pixels}) from the two chips were aligned and sent to the oscilloscope with the largest sampling rate to make Time-Of-Flight (TOF) measurements. The data from pixels OA1 and OA2 of the two DUTs, which were not aligned among the two DUTs, were also acquired.

\begin{figure}[!htb]
\centering\includegraphics[width=.80\textwidth,trim=10 10 10 10]{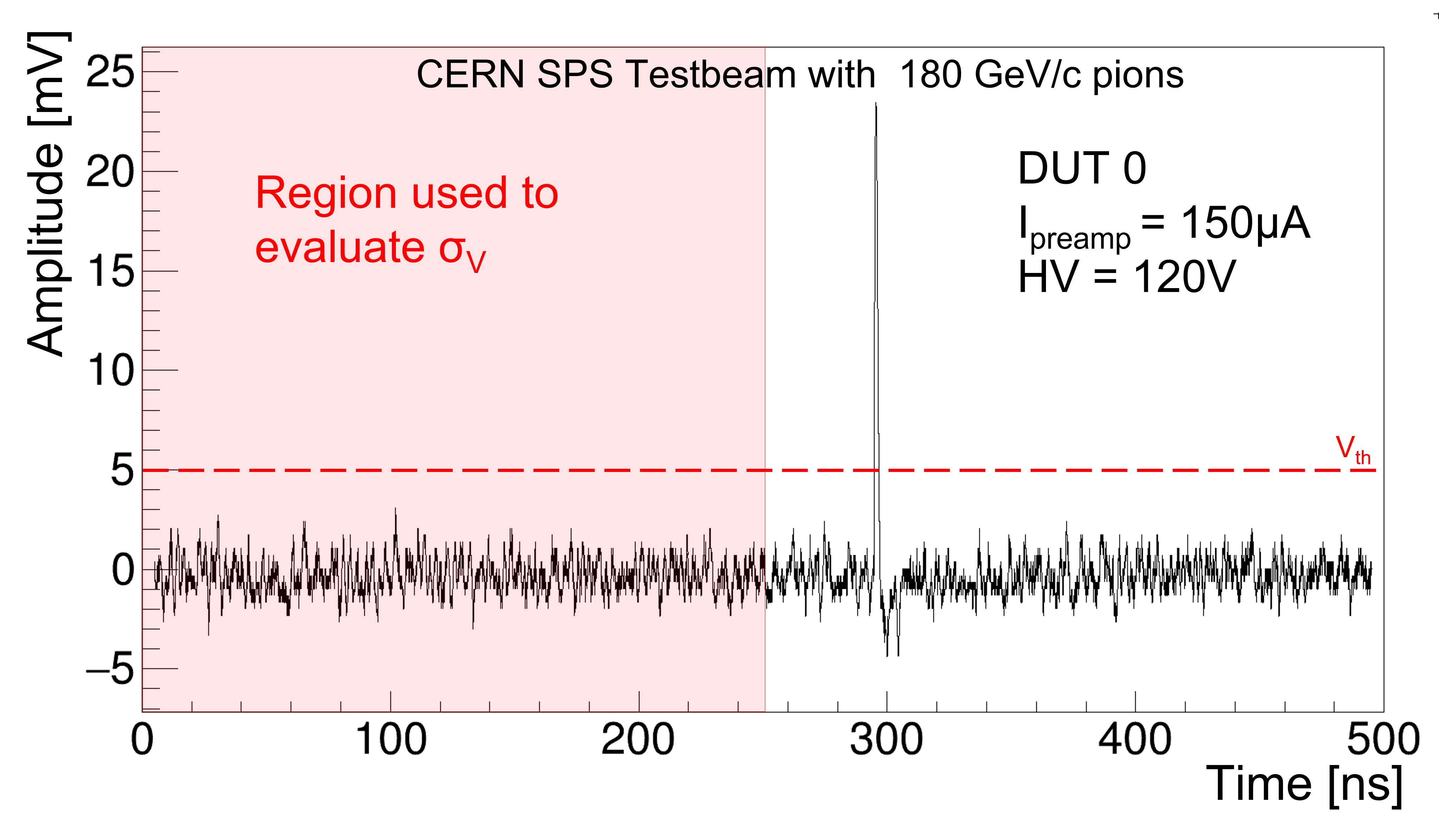}
\caption{\label{fig:waveform} Example of a waveform from working point $ \ipreamp = \SI{150}{\micro\ampere} $. The shaded region below 250 ns is the portion of the waveform used to extract $\sigma_{V}$. The dashed line shows the discrimination threshold used for this working point.
}
\end{figure}

The analysis of the data was performed using the full waveform information acquired by the oscilloscopes. The signals from the DUTs were delayed to guarantee that they were always in the second half of the waveform time window acquired by the oscilloscope.
This configuration allowed using the first half of the waveform to determine the voltage noise $\sigma_V$ at the output of the analog front-end and set a discrimination threshold $ V_{th} $ as a multiple of the voltage noise, independently for the two DUTs. Figure~\ref{fig:waveform} shows a typical waveform with a signal pulse from a MIP at the working point with  $ \ipreamp = \SI{150}{\micro\ampere} $. The dashed line represents the discrimination threshold at $V_{th}=6~\sigma_{V}$.

The FEI4 telescope provided the trigger to the oscilloscopes. A Region Of Interest (ROI) of $ 250 \times 600 $  \si{\micro\meter\squared} was set on one of the trigger planes of the telescope, centered around the pixels OA0 of the two DUTs that were aligned with respect to the beamline and used for the TOF measurement.

Data were acquired 
with DUT0 and DUT1
at the same four front-end working points used for the characterization of the DUTs  with radioactive sources (Section \ref{sec:calibrations}) for a bias voltage of 120 V. At this potential the substrate is not fully depleted. The depletion depth is estimated to be $ \SI{23}{\micro\meter} $, which corresponds to a most probable charge $ Q_{MPV} \approx 1300 $ electrons for a MIP. 

A high voltage scan was also performed  only for the working point $\ipreamp = $ \SI{150}{\micro\ampere}.

To evaluate  the efficiency and time resolution of our DUTs in the cleanest possible way, a selection was applied on the quality of the tracks reconstructed by the FEI4 telescope.
The selection consisted in discarding  events in which more than one track was reconstructed by the telescope, and in accepting only those events with the reconstructed track having an associated hit  in each of the five telescope planes and a  $ \chi^{2}/NDF \le 1 $. 
About 30\% of the triggered events survived this stringent selection on the tracks from the telescope.
%
%
%
%
For this final sample the telescope pointing resolution on the DUT planes was estimated to be  approximately 10 $\mu$m \cite{mateus_thesis}.

\subsection{Cross talk and robustness to induced noise}
\label{subsec:crosstalk}
During laboratory and testbeam measurements,  cross talk between the channels was observed for  events with  large charge deposition,
corresponding to approximately five times the MIP most probable charge. The analysis of these events showed that this cross talk was not influenced by the relative position of the pixels within the matrix or the routing of the signal before and after the amplification. Therefore the observed cross talk could be ascribed to two possible causes. The first is a feedback path passing through the ground of the board, which is then injected to the backside of the chip through the HV decoupling capacitors and finally in the amplifier input. This hypothesis is supported by the fact that the system becomes less stable at bias voltages below \SI{50}{\volt}, when the pixel capacitance is significantly larger. The second cause is noise induced by the driver pulse propagating in the power supply,
since in this prototype the analog and driver electronics share the same supply lines.

As a consequence of this cross talk, only  events in which the pixel under test had the largest signal among the three acquired pixels were selected for the time resolution measurement. No selection was applied for cross talk in the efficiency measurement, since cross talk appears only for efficient events.

\subsection{Efficiency measurement}
For the calculation of the efficiency, all the selected pion tracks reconstructed by the FEI4 telescope were extrapolated to the surface of each of the two DUTs. Only the tracks crossing a DUT within the  area of the three pixels under study, or outside the  external edges of the pixels by at most one standard deviation of the telescope resolution, were retained.
An event was considered efficient if the signal in at least one of the pixels crossed the discrimination threshold.

\begin{figure}[!htb]
\centering
\includegraphics[width=.99\textwidth,trim=0 0 0 0]{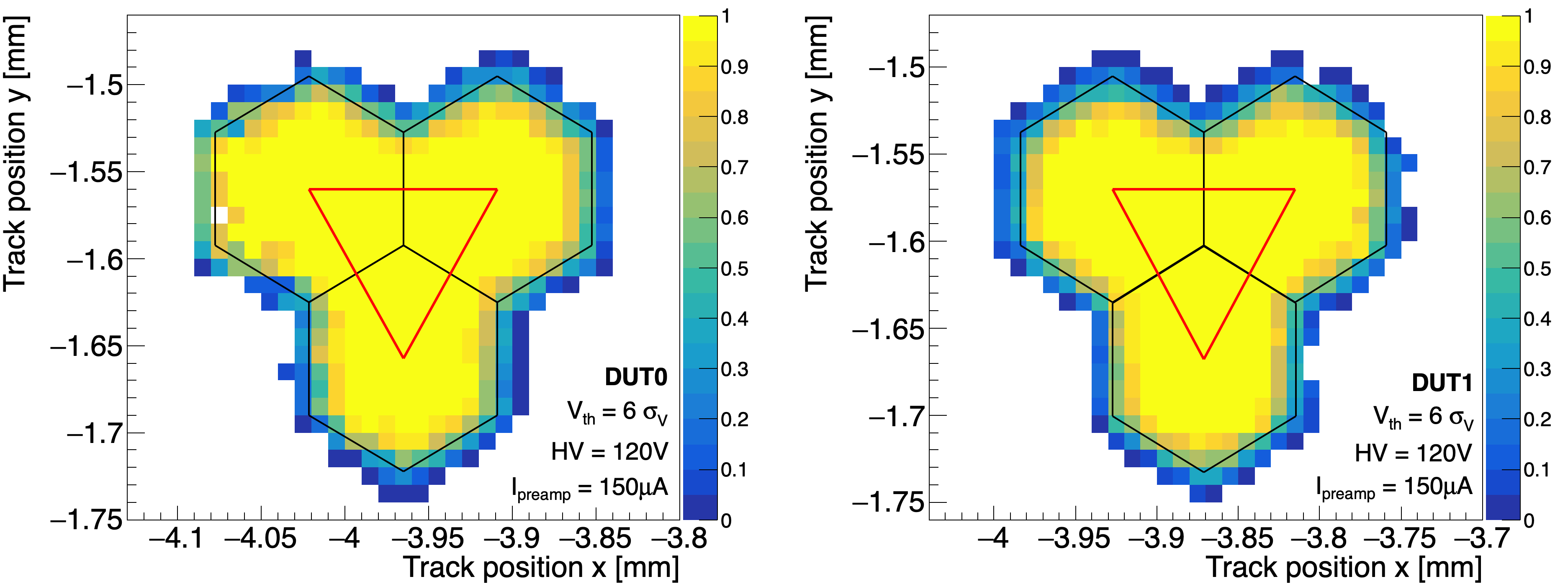}
\caption{\label{fig:effmap} Efficiency map measured for DUT0 (left) and DUT1 (right) at $ \ipreamp = \SI{150} \mu$A, threshold $ V_{th} = 6 ~\sigma_{V}$ and $ HV = \SI{120}{\volt} $. 
The pixel edges are shown by the black lines.
The  efficiency degradation around the external edges of the three pixels is due to the FEI4 telescope resolution.
The efficiency measured inside the triangular area delimited by the red lines is unaffected by the telescope pointing resolution and is used throughout this study.
}
\end{figure}

Figure \ref{fig:effmap} shows the efficiency map for the two DUTs at the $\ipreamp =$ \SI{150}{\micro\ampere} working point, for a threshold of $ 6\sigma_{V} $ and a sensor bias $HV = $ 120 V. 
The panels show the efficiency map for the entire surface of the three pixels (the pixel edges are represented by the black lines). The degraded efficiency measured nearby the twelve external sides of the hexagonal pixels is produced by the FEI4 telescope resolution, as attested by the fact that such degradation of the efficiency is not observed in the region of the three internal hexagon sides between the  pixels.
To avoid biases stemming from the pointing resolution of the FEI4 telescope, the measurement of the detection efficiency was restricted to events with tracks extrapolated inside
the triangular area\footnote{It should be noted that this triangular area constitutes exactly one sixth of the total area of the three pixels, which is representative of all the zones of a pixel (central-pixel zone; zone in between two pixels; zone in between three pixels) in the right geometrical proportions. Therefore, the efficiency measured inside it provides a reliable estimation of the efficiency over the entire hexagonal pixel area, unbiased by the telescope resolution.} 
in between the three pixels that is delimited by the red lines in Figure \ref{fig:effmap}. 


\begin{figure}[!htb]
\centering 
\includegraphics[width=.7\textwidth,trim=0 0 0 0]{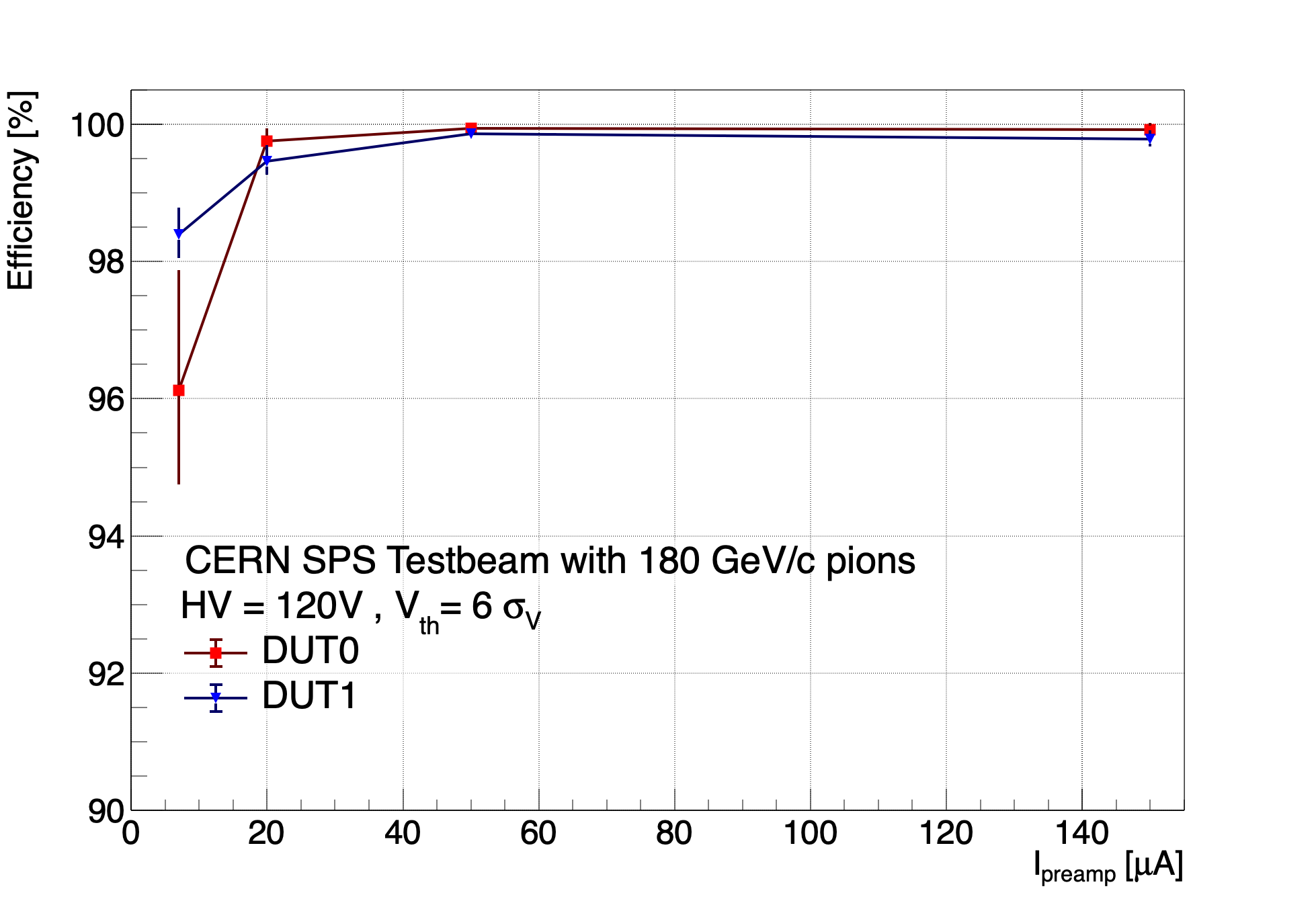}
\caption{\label{fig:effipreampscan} Efficiency vs. $ \ipreamp $ for $ HV = \SI{120}{\volt} $, evaluated in the triangular inter-pixel zone for  DUT0 (in red) and DUT1 (in blue).}
\end{figure}

\begin{figure}[!htb]
\centering 
\includegraphics[width=.65\textwidth,trim=0 0 0 0]{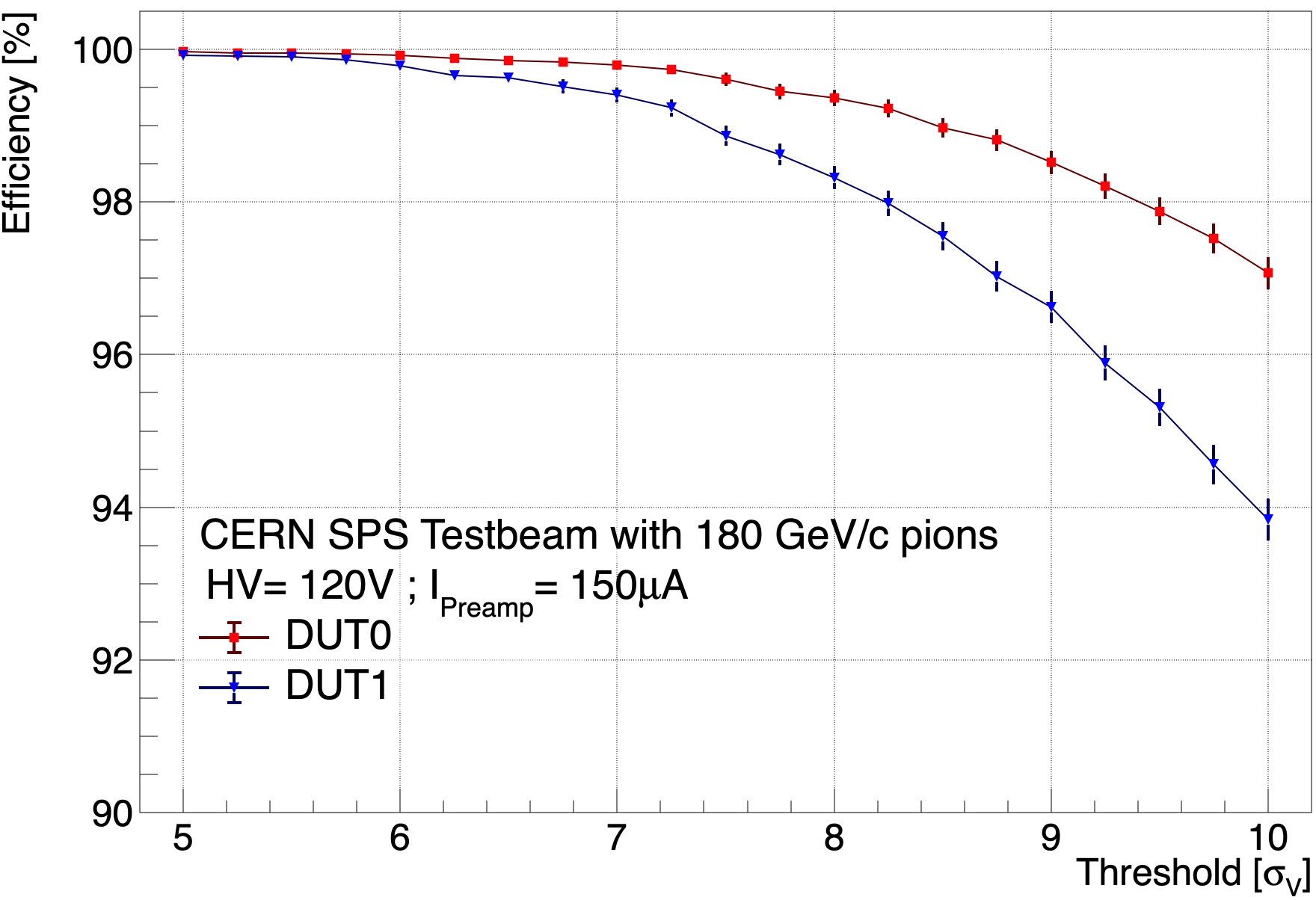}
\caption{\label{fig:effthrscan} Efficiency vs. discrimination threshold measured within the triangular inter-pixel zone for  DUT0 (in red) and DUT1 (in blue).}
\end{figure}

Figure~\ref{fig:effipreampscan} and Table \ref{tab:efftable} show the efficiency obtained within the triangular  area for the four preamplifier working points at a discrimination threshold of 6 $\sigma_V$ and at a sensor bias voltage of 120 V. The efficiency follows the trend expected from the ENC measurement of Table~\ref{tab:gaintable}.  All the preamplifier working points show an efficiency well above 99\% except for the one at \SI{7}{\micro\ampere}. At this current a larger depletion depth would be required to operate the front-end at even higher efficiency. It is also noted that DUT0 shows a lower efficiency, compatibly with the larger ENC measured with the $ \Cd $ source (Section~\ref{sec:calibrations}). At higher currents  DUT0 is slightly more efficient than  DUT1. To investigate this difference, the efficiency as a function of the discrimination threshold was inspected.
The results are reported in Figure \ref{fig:effthrscan} for $ HV = \SI{120}{\volt} $.
The threshold scan indicates a clear difference in performance, in spite of the fact that both DUTs reach the efficiency plateau at a threshold of six standard deviations of the noise. 

\begin{table}[!htb]
\centering
\renewcommand{\arraystretch}{1.3}
\begin{tabular}{c|cccc|l}
\cline{1-5}
\multicolumn{5}{|c|}{Efficiency measured at HV = 120 V}                                                                                                                         & \multicolumn{1}{c}{\textbf{}} \\ \cline{1-5}
\multicolumn{1}{|c|}{$\ipreamp$ [$\mu$A]} & \multicolumn{1}{c|}{7} & \multicolumn{1}{c|}{20} & \multicolumn{1}{c|}{50} & 150 &                               \\ \cline{1-5}
\multicolumn{1}{|c|}{Efficiency DUT0 {[}\%{]}} & \multicolumn{1}{c|}{$ 96.1_{-1.7}^{+1.4} $} & \multicolumn{1}{c|}{$ 99.75_{-0.17}^{+0.12} $} & \multicolumn{1}{c|}{$ 99.94_{-0.05}^{+0.03} $} & $ 99.91_{-0.08}^{+0.05} $ &                               \\ \cline{1-5}
\multicolumn{1}{|c|}{Efficiency DUT1 {[}\%{]}} & \multicolumn{1}{c|}{$ 98.4_{-0.4}^{+0.3} $} & \multicolumn{1}{c|}{$ 99.45_{-0.2}^{+0.2} $}   & \multicolumn{1}{c|}{$ 99.86_{-0.07}^{+0.05} $} & $ 99.78_{-0.11}^{+0.08} $ &                               \\ \cline{1-5}
\end{tabular}
\caption{Efficiency of the two DUTs at different $\ipreamp$ for $ HV = \SI{120}{\volt}$. The efficiency is measured according to the definition given in the text. The uncertainties are statistical only.}
\label{tab:efftable}
\end{table}

\begin{figure}[!htb]
\centering 
\includegraphics[width=.65\textwidth,trim=0 0 0 0]{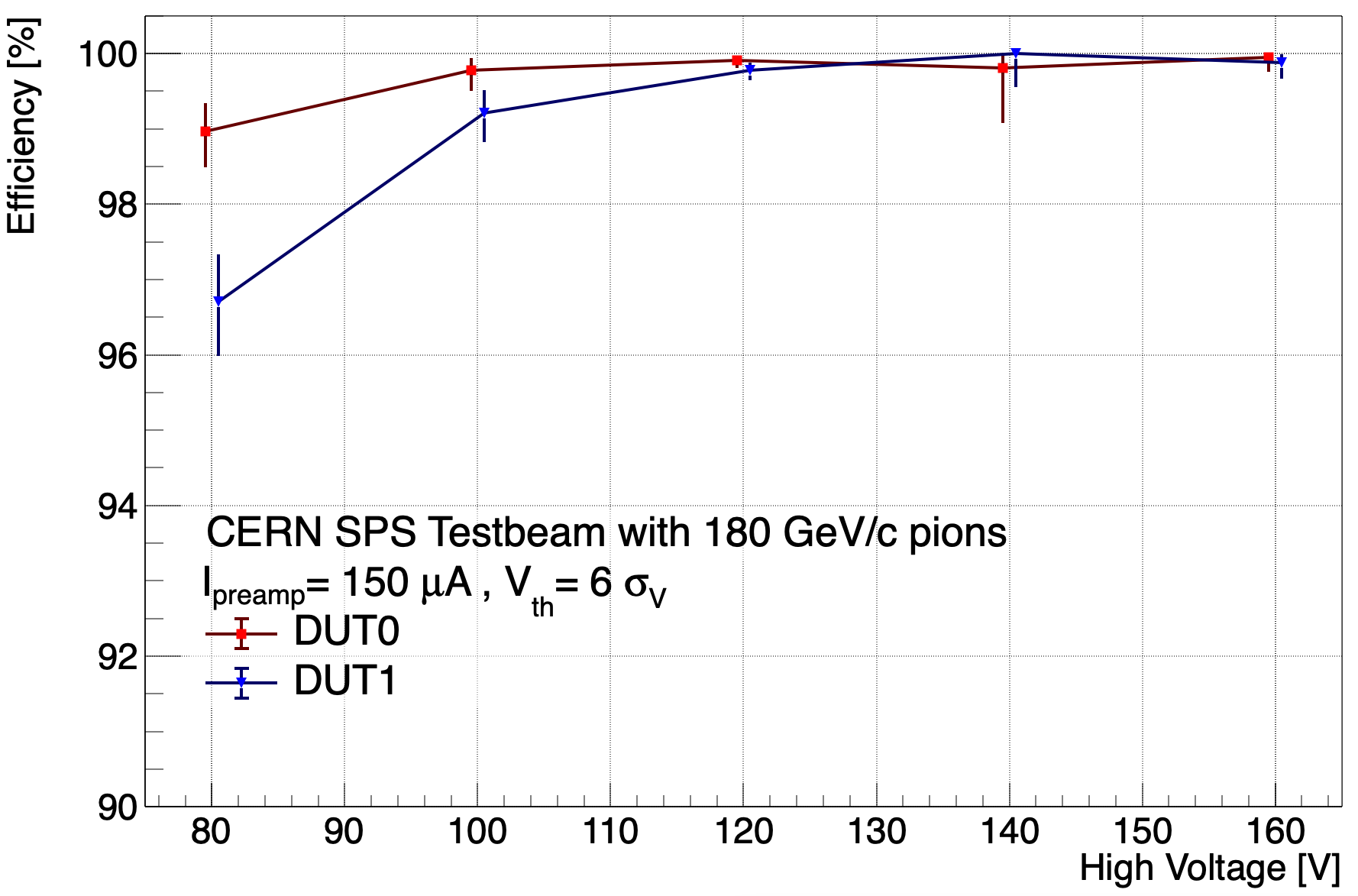}
\caption{\label{fig:effHVscan} Efficiency vs. \!\!sensor bias voltage for the two DUTs at $\ipreamp$ = 150 $\mu$A and voltage threshold of 6 $\sigma_V$. The vertical error bars show the statistical uncertainties.}
\end{figure}

\begin{table}[!htb]
\centering
\renewcommand{\arraystretch}{1.3}
\begin{tabular}{c|ccccc|l}
\cline{1-6}
\multicolumn{6}{|c|}{Efficiency measured at $\ipreamp = 150\mu A$}                                                                                                                         & \multicolumn{1}{c}{\textbf{}} \\ \cline{1-6}
\multicolumn{1}{|c|}{HV [V]} & \multicolumn{1}{c|}{80} & \multicolumn{1}{c|}{100} & \multicolumn{1}{c|}{120} &\multicolumn{1}{c|}{140} & 160 &                               \\ \cline{1-6}
\multicolumn{1}{|c|}{Efficiency DUT0 {[}\%{]}} & \multicolumn{1}{c|}{$ 98.97_{-0.46}^{+0.35} $} & \multicolumn{1}{c|}{$ 99.78_{-0.26}^{+0.14} $} & \multicolumn{1}{c|}{$ 99.91_{-0.08}^{+0.05} $} &\multicolumn{1}{c|}{$ 99.81_{-0.71}^{+0.18} $} & $ 99.95_{-0.18}^{+0.04} $ &                               \\ \cline{1-6}
\multicolumn{1}{|c|}{Efficiency DUT1 {[}\%{]}} & \multicolumn{1}{c|}{$ 96.70_{-0.70}^{+0.61} $} & \multicolumn{1}{c|}{$ 99.21_{-0.37}^{+0.28} $}   & \multicolumn{1}{c|}{$ 99.78_{-0.11}^{+0.08} $} &\multicolumn{1}{c|}{$ 100._{-0.42}^{+0.00} $} & $ 99.88_{-0.20}^{+0.09} $ &                               \\ \cline{1-6}
\end{tabular}
\caption{Efficiency at different HV values for the two DUTs for $\ipreamp$ = 150 µA. The efficiency is measured according to the definition given in the text. The uncertainties are statistical only.}
\label{tab:efftablehv}
\end{table}

The efficiency measurement of a sensor HV scan carried out for the working point  $ \ipreamp = \SI{150}{\micro\ampere} $ is reported in Figure~\ref{fig:effHVscan} and Table~\ref{tab:efftablehv}. 
The scan shows that the two DUTs are in the efficiency plateau at 120 V. At the \SI{50}{\ohm\cm} substrate resistivity of this prototype, increasing the HV from \SI{120}{\volt} to \SI{160}{\volt} increases the depletion depth by 15\%, which has little or no contribution for the working point at the highest power consumption, but  could have been beneficial for the front-end operation at \SI{7}{\micro\ampere} for which a small drop in efficiency was observed (Figure~\ref{fig:effipreampscan}).

\subsection{Time resolution measurement}
For the time resolution measurement, the pixels OA0 of the two DUTs were carefully aligned with respect to the beamline and events in which the two pixels registered signals with amplitudes above a 
discrimination threshold of $ 6~\sigma_V $ in coincidence were selected.
Furthermore, the telescope-track quality selection described in Section~\ref{subsec:dataset} and the cross-talk selection described in Section~\ref{subsec:crosstalk} were applied. To avoid biasing the sample with a geometrical selection, no requirement on the telescope-track position was imposed.

\begin{figure}[!htb]
\centering %
\includegraphics[width=.92\textwidth,trim=5 0 670 0, clip]{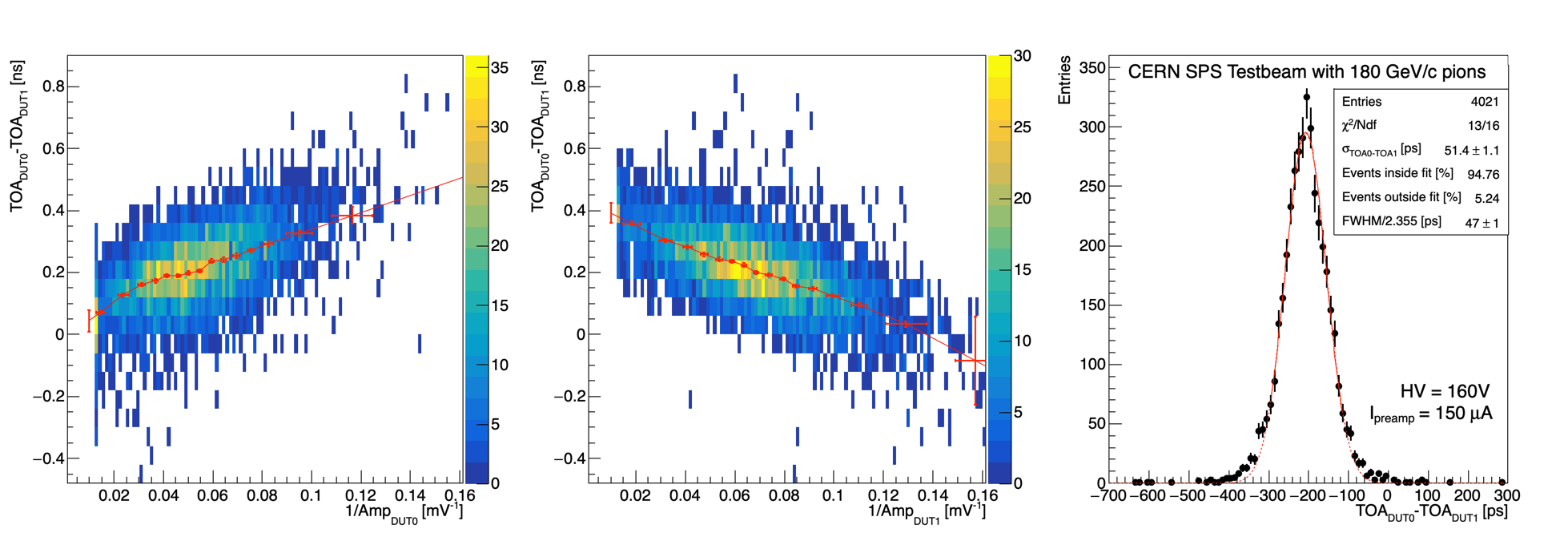}
\caption{\label{fig:TWcorr} Distributions of the difference in TOA between the two DUTs vs. the inverse of the amplitude that was used for the time-walk correction of DUT0 (left) and DUT1 (right). Both DUTs were operated at $ \ipreamp = \SI{150}{\micro\ampere} $ and $ HV = \SI{160}{\volt} $. The time-walk correction points (in red) were obtained by  a Gaussian fit on each bin of the inverse of the amplitude. The red segments show the  linear interpolation between the time-walk correction points used to correct the data.
The TOA difference contains an arbitrary offset that is irrelevant for the measurement of the time resolution.}
\end{figure}

{\it Time-walk correction} 

Figure \ref{fig:TWcorr} shows the 
difference in the Time-Of-Arrival (TOA) measured in pixels OA0 of  DUT0 (TOA0) and DUT1 (TOA1) as a function of the inverse of the signal amplitude in DUT0 (left) and DUT1 (right)
for the working point at \SI{150}{\micro\ampere} and $ HV = \SI{160}{\volt}$.
The data show a large variation of the average of the difference  TOA0$-$TOA1  as a function of the signal amplitudes, of the order of a few hundreds ps, that was corrected in the following way. 

The data were divided in variable-size bins of the inverse of the amplitude containing at least 200 entries. For each of these bins for DUT0 in Figure \ref{fig:TWcorr} left, the most probable value of TOA0-TOA1 was obtained by  a Gaussian fit (red points in the figure). That value was associated to the average value of the inverse of the DUT0 amplitude distribution within that bin (instead than to the center of the bin).
An event-by-event correction was then applied to the inverse of the amplitude of the  signal in DUT0, using the value provided by the linear interpolation (red segments in the figure) of the two adjacent  time-walk correction points.
Once DUT0 was time-walk corrected in this way, the entire procedure was repeated for DUT1 (shown in Figure \ref{fig:TWcorr} right) to complete the time-walk correction\footnote{Given its importance for this measurement, the time-walk correction was performed also with an unbinned maximum-likelihood fit. This second method was used for a simultaneous extraction of the  resolution parameter \sigtoa and of the two  time-walk correction functions for DUT0 and DUT1. For all the data samples analysed, the results  were within few percent from those obtained by the method described in the text.}.

{\it Extraction of the time resolution} 

Once  data were corrected for time walk,
Gaussian fits were performed to the TOA0-TOA1 distributions,  including only  bins containing more than 25\% of the entries at the maximum  of the distribution. It was then assumed that the two DUTs have the same resolution, so that the time resolution of each DUT can be estimated as $\sigma_{t} = \sigma\_{\it TOA0-TOA1}/\sqrt{2}$.

\begin{figure}[!htb]
\centering %
\includegraphics[width=.47\textwidth,trim=1315 0 40 60, clip]{./Figures/TimeRes_WP7_160V.png}
\qquad
\includegraphics[width=.47\textwidth,trim=1315 0 40 60, clip]{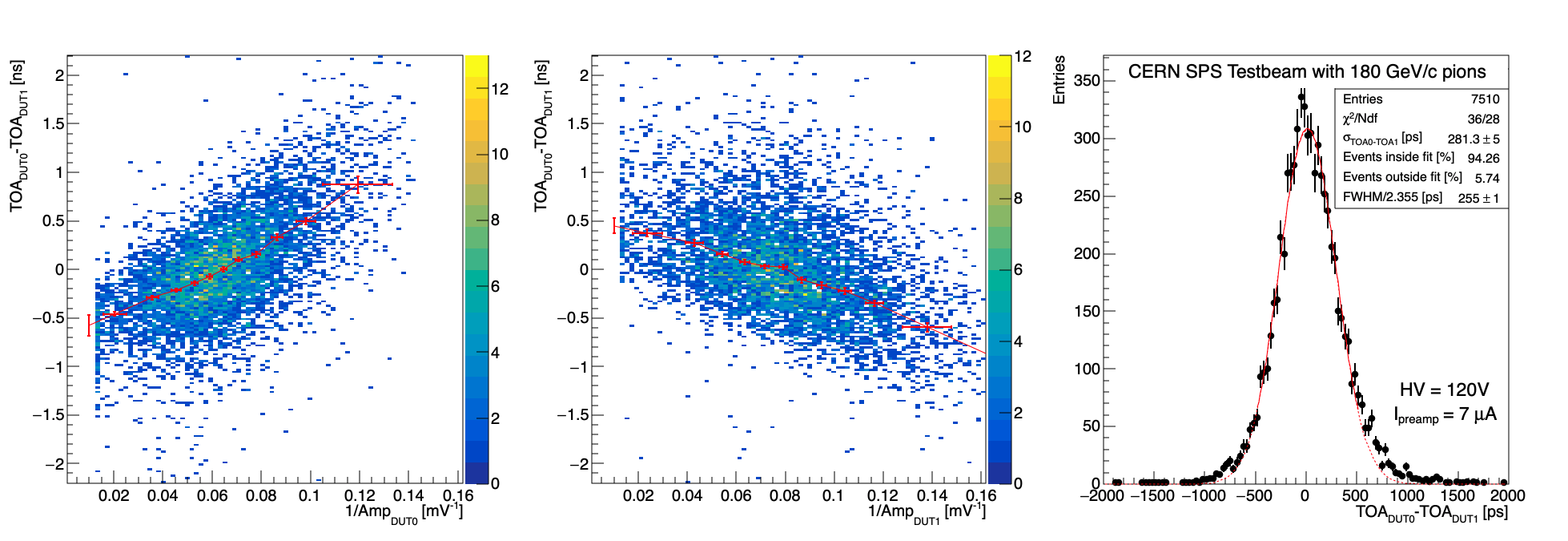}
\caption{\label{fig:TOF} TOA difference between pixels OA0 of DUT0 and DUT1 after time-walk correction for the two working points reported in the panels. 
A constant arbitrary offset is present, which is irrelevant for the time-resolution calculation. The  red lines show the results of the Gaussian fit using only the bins with more than 25\% of the entries in the maximum of the distribution. 
The full red lines show the ranges used for the fits, while the dashed red lines allow the estimation of the non-Gaussian components in the tails.}
\end{figure}

As an example, Figure \ref{fig:TOF} shows the resulting TOA difference distribution after time-walk correction for the data acquired at the working points  \SI{150}{\micro\ampere} and $ HV = \SI{160}{\volt}$ (left) and 7 $\mu$A and 120 V (right).  In the case of the former, that is the  best working point for time resolution, the standard deviation obtained by the Gaussian fit is measured to be $ \sigma\_{\it TOA0-TOA1}= (51.4 \pm 1.1) $ ps. 
Therefore the time resolution of each DUT is estimated to be
\begin{equation}
    \sigma_{t} = \frac{\sigma\_{\it TOA0-TOA1}}{\sqrt{2}} = (36.4 \pm 0.8) \si{\pico\second}.
\end{equation}
The fraction of events exceeding the Gaussian fit in the tails of the distribution of Figure \ref{fig:TOF} is approximately 5\%. This fraction of events represents the typical non-Gaussian component found in the tails of the time-resolution distribution for  all the data sets acquired at the testbeam at different sensor and preamplifier bias.
As a consequence the resolutions quoted in the following refer to 95\% of the signals acquired by the DUTs.

\begin{figure}[!htb]
\centering %
\includegraphics[width=.65\textwidth,trim=0 0 0 0, clip]{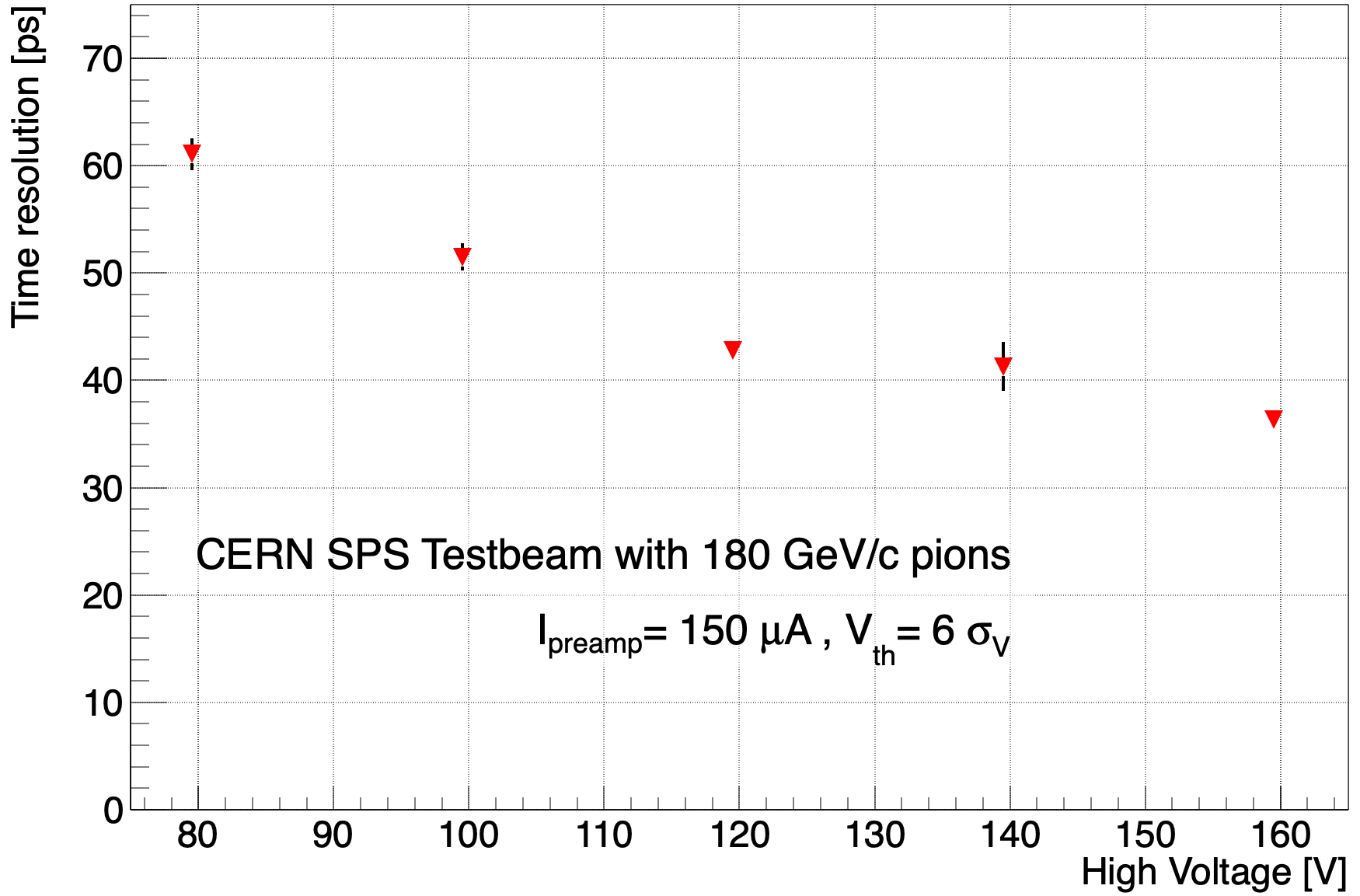}

\vspace{10pt}
\includegraphics[width=.65\textwidth,trim=0 0 0 0, clip]{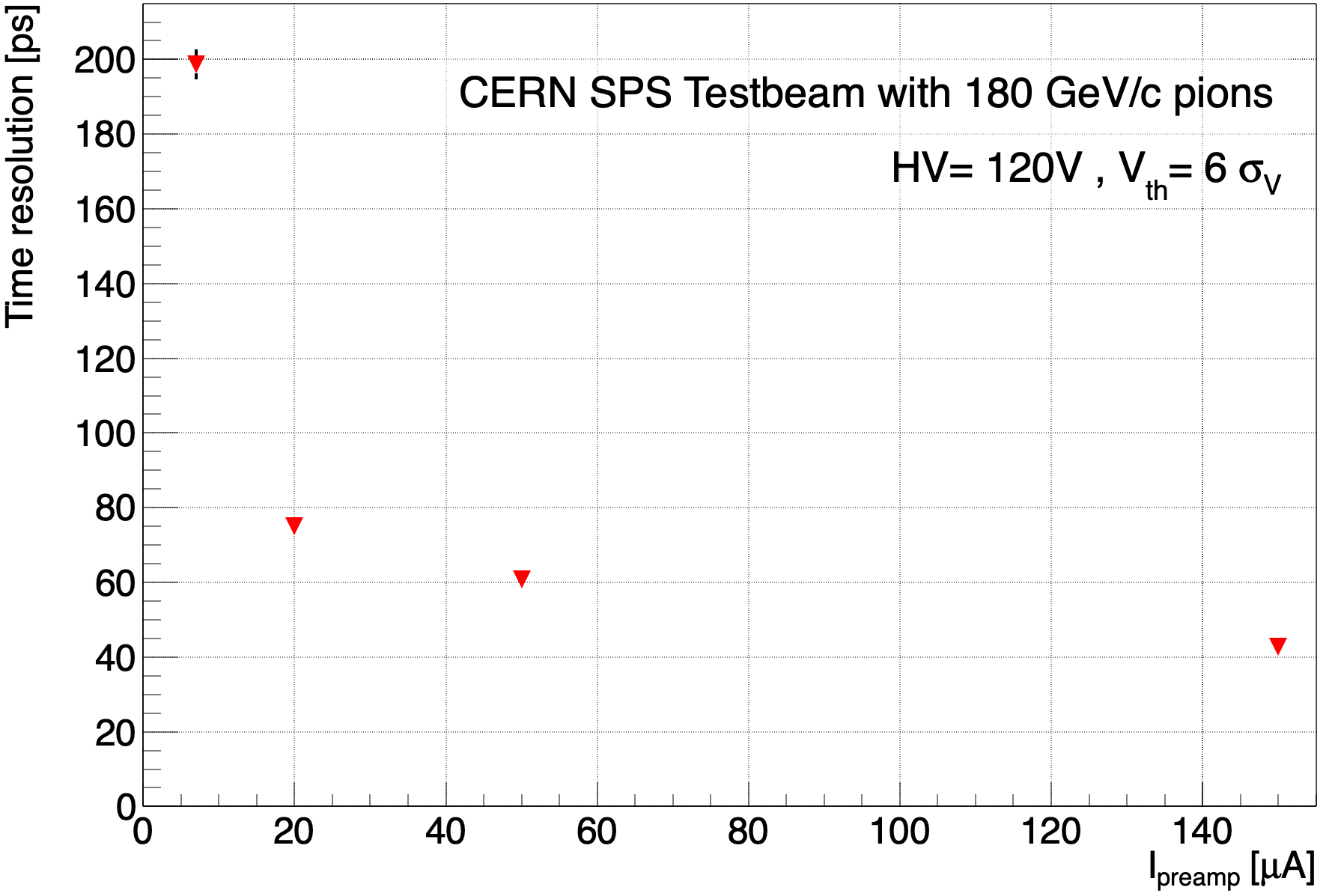}
\caption{\label{fig:TOFHV} Top: time resolution as a function of sensor bias voltage at $ \ipreamp = \SI{150}{\micro\ampere}$. 
Bottom: time resolution as a function of $ \ipreamp $ for sensor bias voltage $HV = $ 120 $V$.
The time resolution is defined as $(\sigma\_{\it TOA0-TOA1})/\sqrt{2}$. It refers to the Gaussian component of the data, which is approximately 95\% of the total.}
\end{figure}

Figure \ref{fig:TOFHV} top shows the time resolution as a function of the HV for the highest power consumption working point   $ \ipreamp = \SI{150}{\micro\ampere} $. 
The time resolution varies between 60 and 36 ps with the HV between 80 and 160 V.
At $ HV = \SI{120}{\volt} $ the timing performance is approximately 20\% worse than the one measured at 160 V.

Figure \ref{fig:TOFHV} bottom shows the time resolution measured at $ HV = \SI{120}{\volt} $  for the  four $\ipreamp$ working points. As expected, the time resolution depends on the preamplifier current. A significant degradation of the performance is observed for the lowest power-consumption working point studied $\ipreamp = $  \SI{7}{\micro\ampere}, for which the time resolution still remains  at the level of 200 ps.

\vspace{5pt}


\section{Conclusions}
\label{sec:discussion}

A monolithic silicon  detector prototype with 100 µm pixel pitch was produced in 130 nm SiGe BiCMOS technology on \SI{50}{\ohm\centi\meter} wafers and tested at a beamline with minimum-ionizing particles. The analysis of the data acquired with analog channels using a threshold of $6~\sigma_{V}$ shows that this detector technology can achieve efficiencies up to 99.9\%.
This result proves that the low ENC reachable with SiGe HBTs allows operating at very high efficiency even with a thin depletion layer of approximately 20 µm. The small drop in efficiency observed at the lowest amplifier current studied, caused by an increase of the ENC, could be  compensated by increasing the sensor depletion depth. Therefore, further optimization of this detector technology is possible by adopting higher resistivity substrates to obtain full depletion of the sensitive volume.

Results with a previous prototype~\cite{Paolozzi_2020} that provided uniquely the TOT measurement to perform the time-walk correction, showed that a remarkable timing performance at the level of 50 ps was possible only at large discrimination threshold, therefore compromising the detection efficiency. 
The data presented here demonstrate that a time-walk correction that uses the signal amplitude instead of the TOT can provide an even better timing performance at the lowest discrimination threshold permitted by the ENC of the amplifier:
when operated at a preamplifier current as low as \SI{20}{\micro\ampere} the SiGe HBT amplifier implemented in the ASIC provides a time resolution better than \SI{100}{\pico\second} and an efficiency of 99.6\%;
at a preamplifier current of \SI{150}{\micro\ampere} a time resolution of \SI{36}{\pico\second} and an efficiency of 99.9\% are measured, even without the support of an avalanche gain layer to boost the signal-to-noise ratio. These results prove that this technology is suitable for applications that require the combination of tracking capabilities with excellent time resolution.

\acknowledgments
The authors wish to thank Alexander Gebershagen and the CERN SPS H8-beamline  team for the  support and coordination during the experimental activities, as well as the technical staff of the DPNC of the University of Geneva. 
This research would not have been possible without the support of the H2020 programme through the ERC Advanced MONOLITH project (grant agreement ID 884447) and the ATTRACT Phase1 MonPicoAD project, as well as of the Swiss National Science Foundation grant number 200020\_188489.



\newpage
\bibliographystyle{unsrt}
\bibliography{bibliography.bib}

\end{document}